\documentclass [11pt,a4paper]{article}

\usepackage{shrthnds}
\usepackage{cite}
\usepackage{graphicx}
\usepackage{color}
\usepackage{subfigure}

\usepackage{multirow}

\usepackage[hang,small]{caption}

\usepackage{geometry}
\usepackage{titlesec,titletoc}
  \titleformat{\section}{\Large\bf}{\thesection.}{1em}{}
  \titleformat{\subsection}{\large\bf}{\thesubsection.}{1em}{}

\title{\bf Standard Model with Cosmologically Broken Quantum Scale Invariance}

\author{\bf Pankaj Jain$^{1}$\footnote{email: pkjain@iitk.ac.in}~and~Subhadip Mitra$^{2}$\footnote{email: subhadip@imsc.res.in} }

\date{10 March, 2009}%{\today}

\begin{document}
 \maketitle
\vspace{-0.6cm}
\bc
{\small 1) Department of Physics, IIT Kanpur, \\Kanpur 208 016, India\\
2) The Institute of Mathematical Sciences, \\Chennai 600 113, India}
\ec

\centerline{\date{10 March, 2009}}
\vspace{0.5cm}

\bc
\begin{minipage}{0.8\textwidth}{\small {\bf Abstract:}
We argue that scale invariance is not anomalous in quantum field theory,
provided it is broken cosmologically. We consider a locally scale invariant
extension of the Standard Model of particle physics and argue that
it fits both the particle and cosmological observations. The model
is scale invariant both classically and quantum mechanically. The scale
invariance is broken cosmologically producing all the dimensionful
parameters. The cosmological constant or dark energy is a prediction
of the theory and can be calculated systematically order by order in perturbation
theory. It is expected to be finite at all orders. The model does
not suffer from the hierarchy problem due to absence of scalar particles,
including the Higgs, from the physical spectrum.
}\end{minipage}
\ec

\vspace{0.5cm}

\section{Introduction}
In a recent paper \cite{JMS} we have considered the possibility 
that scale invariance may
be an exact symmetry in quantum field theory. The basic idea makes nontrivial
use of the concept of cosmological symmetry breaking \cite{JM}.
The possibility that scale invariance may be an exact symmetry in quantum
field theory has been considered earlier by many authors 
\cite{Kallosh,Englert,Fradkin78,Shore,Fradkin82,Buchbinder85,Odintsov91,Odintsov92,Odintsov93,Buchbinder,Shaposhnikov1,Shaposhnikov2}.  
An interesting proposal for implementing scale invariance in Standard Model 
has been 
introduced by Cheng and collaborators \cite{ChengPRL,ChengKao}.
In Refs. \cite{ChengPRL,ChengKao} the authors proposed a locally pseudo-scale
invariant Standard Model by introducing the Weyl vector meson.
In Refs. \cite{ChengPRL,ChengKao} the authors split the scale transformation
into a pseudo-scale transformation and general coordinate
invariance. The pseudo-scale transformation is similar to the conformal
transformations studied earlier \cite{Englert}.
Hence if the action satisfies both pseudo-scale and general
coordinate invariance it also obeys scale invariance.
The Higgs particle is interpreted as the longitudinal mode of the Weyl
meson and hence disappears from the particle spectrum.
Phenomenological consequences of this model have also been
studied in Refs. \cite{Huang,Wei,JMS,AJS}. 
Local scale invariance has also been studied
in Refs. \cite{Padmanabhan85,Hochberg,Wood,Wheeler,Feoli,Pawlowski,Nishino,Demir}.

The phenomenon of cosmological symmetry breaking
is inspired by the standard big bang model. Here,
at the leading order,
the universe is described by a homogeneous and isotropic
Friedmann-Robertson-Walker (FRW) metric. The essential point is that
the universe is a time dependent solution of the
classical equations of motion. All
physical phenomena take place in this background.
It is not just the metric which might be time dependent.
If we add some fields to the action
besides gravity, then these fields may also acquire time dependence,
as, for example, happens in the case of slow roll models of dark energy
\cite{Wetterich,Ratra,Fujii,Chiba,Carroll1,Caldwell,Uzan,Amendola,Bean}.
Hence to study any process
we need to make a quantum expansion
around this classical time dependent solution. The important point is that
these classical fields need not take values which minimize the potential.
It was shown in Ref. \cite{JM} that an expansion around such a background
time dependent solution breaks some of the symmetries of the action. It
was further argued \cite{JM,JMS} that this is a particularly attractive
way to break scale invariance. This phenomenon has most of the attractive
features of the well known mechanism of spontaneous symmetry breaking. 
However the actual mechanism is very different. For example, in contrast
to spontaneous symmetry breaking, we do not predict any zero mass Goldstone
boson if the symmetry is broken cosmologically. The phenomenon of cosmological
symmetry breaking is naturally implemented in the scale invariant 
Standard Model \cite{ChengPRL}. In this case, one finds that this model 
leads to both dark energy and dark matter \cite{JMS,AJS}. 
Within the framework of global scale invariance the possibility that 
the background curvature can lead to symmetry breaking has also been considered
earlier 
\cite{Shore,Cooper,Allen,Ishikawa,Wetterich1,Buchbinder85,Perez,Finelli,Wetterich2,Buchbinder}.

A fundamental problem with imposing scale invariance is that it may
be anomalous \cite{Coleman,Collins,Fujikawa}. Hence it is not clear whether
the locally pseudo-scale invariant model proposed in Ref. \cite{ChengPRL} is
meaningful quantum mechanically. In Ref. \cite {Kallosh} it was conjectured
that conformal invariance may not be anomalous.
In Ref. \cite{Englert}, the authors showed
conformal invariance is not anomalous if it is suitably extended to
arbitrary number of dimensions, provided the symmetry is
broken spontaneously.
We have also argued in Ref. \cite{JMS} that
pseudo-scale invariance need not be
anomalous in theories which are cosmologically broken.
Here we study scale invariance in the context of cosmological
symmetry breaking. We also apply this to the
Standard Model of particle physics with local scale invariance \cite{ChengPRL}.
An interesting
prediction of pseudo-scale invariance is that it does not
admit a cosmological constant
term in the action \cite{JMS,Wetterich,Wetterich2,JM09}.
Hence this symmetry might potentially solve the cosmological constant problem
\cite{Weinberg,Peebles,Padmanabhan,Copeland,Carroll,Sahni,Ellis}.
However cosmological constant is generated due
to the phenomenon of cosmological symmetry breaking \cite{JM,JMS}.
If pseudo-scale invariance indeed holds at the quantum level then we expect that
we should find finite results for cosmological constant at one loop.
In a recent paper \cite{JM09} we have explicitly demonstrated this result
and computed the finite value of the one loop contribution to the
cosmological constant. The calculation in Ref. \cite{JM09} was performed
in the adiabatic limit, where we assume that the background classical
solution is very slowly varying with time. 
There also exist several other approaches to solving the cosmological
constant problem \cite{Weinberg,Aurilia,VanDer,Henneaux,Brown,Buchmuller,Henneaux89,Sorkin,Sundrum}.
The possibility that scale or conformal invariance might provide a solution
to the cosmological constant problem has also been discussed in
\cite{Wetterich,Wetterich2}. Here the author works within the framework
of anomalous conformal invariance. The cosmological 
constant problem 
is solved by assuming a fixed point in the beta function. The model we consider
in the present paper is very different since we impose exact local conformal
invariance. 
We demonstrate that a locally conformal invariant extension of the 
Standard Model 
fits both the cosmological and high energy physics observations if the 
conformal invariance is broken cosmologically.

The essential idea can be captured by considering a simple model where
we include only one real scalar field besides gravity.
This simple model displays global scale invariance.
The action
for this model may be written as,
\begin{equation}
\mc{S} = \int d^4x \sqrt{-g}\left[\frac12g^{\mu\nu}\partial_\mu\Phi\partial_\nu\Phi
-\frac{\lambda}{4}\Phi^4 -\frac{\beta}{8}\Phi^2R \right],
\label{eq:Action_Classical}
\end{equation}
where $R$ is the Ricci scalar and $\Phi$ is a real scalar field.
The model has no dimensionful parameter.
In four dimensions
the pseudo-scale or conformal transformation can be written as follows:
\ba
x & \rightarrow & x\, , \cr
\Phi &\rightarrow & \Phi/\Lambda\, ,\cr
g^{\mu\nu} &\rightarrow & g^{\mu\nu}/\Lambda^2.
\label{eq:PSU}
\ea

In Refs. \cite{JM,JMS} the authors assumed an FRW background metric with
scale factor $a(t)$ and the curvature parameter $k=0$. The
model admits a classical solution \cite{JMS},
\begin{equation}
a(t) = a_0 \exp(H_0t)
\label{eq:a(t)}
\end{equation}
with the scalar field,
\begin{equation}
\Phi_0 = \sqrt{{3\beta\over \lambda}} \, H_0 ,
\label{eq:Phi_cl}
\end{equation}
where $H_0$ is the Hubble constant, which is found to be
independent of time in this case. Similar solutions have also been considered
earlier \cite{Cooper,Buchbinder85,Buchbinder,Finelli,Wetterich,Wetterich2}

We expand the scalar field
$\Phi$ around this classical solution,
\begin{equation}
\Phi(x) = \Phi_0 + \phi(x)\, ,
\label{eq:Phi}
\end{equation}
where $\phi(x)$ represent the quantum fluctuations.
Similarly the metric is expanded around its classical solution.

\section{Quantum Scale Invariance}
In this section we demonstrate that we can extend scale invariance as an
exact symmetry in quantum field theory. Using dimensional regularization, we write a regulated action
which is exactly scale invariant. Here
scale transformation is extended to arbitrary dimensions using the 
earlier works \cite{Englert,JMS}. However our 
precise proposal for the regulated action is different. 
We propose the following regulated action in $d=4-\epsilon$ dimensions,
\begin{equation}
\mathcal{S}
= \int d^dx \sqrt{-\bar{g}}\Bigg({1\over 2}{\bar g}^{\mu\nu}
\partial_\mu\Phi\partial_\nu\Phi
-{\lambda\over 4}\Phi^4 (\bar R^2)^{\epsilon/4}
-\frac{\beta}{8}\Phi^2{\bar R}  \Bigg).
\label{eq:Action_d_dim}
\end{equation}
Here we are using the notation of Ref. \cite{tHooft} and denote all quantum
gravity variables, such as $g_{\mu\nu}$, $R$ etc, with a bar. 
This action is invariant under the generalized pseudo-scale or
conformal transformation,
\ba
x & \rightarrow & x\, , \cr
\Phi &\rightarrow & \Phi/\Lambda\, ,\cr
{\bar g}^{\mu\nu} &\rightarrow & {\bar g}^{\mu\nu}/\Lambda^{b(d)}\, ,\cr
{\bar g}_{\mu\nu} &\rightarrow & {\bar g}_{\mu\nu}\Lambda^{b(d)}\, ,\cr
A_\mu &\rightarrow& A_\mu\, ,\cr
\Psi &\rightarrow& \Psi/\Lambda^{c(d)},
\label{eq:pseudo_ddim}
\ea
where $b(d) = 4/(d-2)$ and $c(d) = (d-1)/(d-2)$. Here we have also included the
transformation law for vector and spinor fields, which we will need later. 
The term $(\bar R^2)^{\epsilon/4}$ which multiplies the $\Phi^4$ term 
is treated by expanding around the classical solution. We have \cite{tHooft},
\begin{eqnarray}
{\bar g}_{\mu\nu} &=& g_{\mu\nu} + h_{\mu\nu},\nonumber\\
\bar{R} &=& R + h_{\beta ;\alpha}^{\beta ;\alpha} -
h_{\alpha ;\ \beta}^{\beta ;\alpha} - h^\nu_\alpha R^\alpha_\nu + ...\, ,
\label{eq:barR}
\end{eqnarray}
where $g_{\mu\nu}$ is the classical metric and $R$ the 
classical curvature scalar. Here we shall assume the classical metric
to be the FRW metric. In this case we find that $R=-12H_0^2$.
 In Eq. \ref{eq:barR}, 
``$;$" denotes covariant derivatives.

The transformation, Eq. \ref{eq:pseudo_ddim},
is similar to what was also proposed
in Ref. \cite{Englert}. However our regulated action differs from earlier
proposals \cite{Englert,JMS} since we use the Ricci scalar, rather than the
scalar field, raised to a fractional power to regulate the quartic coupling
term. The advantage of regulating in this manner is that as long as we
neglect quantum gravity effects, the action is manifestly renormalizable. 
When we expand $\bar {R}$ around its classical solution, then
at the leading order 
we simply recover the standard $\Phi^4$ interaction. At higher 
orders, however, the term $(\bar R^2)^{\epsilon/4}$ leads to 
additional interaction vertices, involving  scalar fields and all
possible powers of the gravitational field. This is an expansion in
powers of $\epsilon$ and hence the corresponding couplings go to zero in
four dimensions. These additional vertices make the higher loop analysis of
quantum gravity somewhat more complicated. 
However these contributions are suppressed and as long as we are not 
interested in quantum gravity corrections, they can be ignored. For the
action given in Eq. \ref{eq:Action_d_dim} these corrections are suppressed
by powers of $1/\beta$. In the regularization proposed in \cite{Englert,JMS},
however, we would pick up additional powers of the scalar field in the action.
These would generate additional interaction vertices which make the loop
analysis of the theory more complicated. Furthermore 
renormalizability of the theory is not obvious even if we ignore 
quantum gravity 
contributions. 

The one loop analysis of the model, Eq. \ref{eq:Action_d_dim}, has
been performed in Ref. \cite{JM09}. In Ref. \cite{JM09}, the authors used
a different regularization and hence in the present case we might expect
some additional terms at one loop. These will arise due to the expansion
of the term $(\bar R^2)^{\epsilon/4}$ which generates additional vertices. 
Since the additional vertices that 
get generated in the next to leading order term 
are proportional to $\epsilon$, it is clear that all
additional contributions are finite at one loop. These can contribute only
to the cosmological constant. For all other quantities these give 
contributions suppressed by powers of $1/\beta$. 
The cosmological
constant in the model, Eq. \ref{eq:Action_d_dim}, 
has been explicitly shown to be finite at one loop
\cite{JM09},
as expected from pseudo-scale invariance.

Finally we define the path integral measure.  
The scalar field measure may be written as 
\ba
\Pi_x{\cal D}\left[(- g)^{1/4} (R^2)^{1/4} \Phi(x) 
\right]
\ea
Here we have used the classical curvature scalar in order to scale the measure
proposed in \cite{Fujikawa}
such that it is invariant under the generalized pseudo-scale 
transformations. It is clear that this is possible only if we are 
expanding around a curved background. Alternatively we could use the classical
field $\Phi_0$ to scale the measure, as long as it acquires a non-zero
value. It is clear that both the
action and the measure are exactly invariant under the generalized
pseudo-scale transformations.
Here we specialize only to the scalar field measure. The full quantum
gravity measure may also be constructed following
Refs. \cite{Fujikawa83,Fradkin}.

We emphasize that here we have shown that the theory is exactly invariant
under a generalized pseudo-scale transformation, displayed in Eq. 
\ref{eq:pseudo_ddim}. Hence this symmetry transformation is not anomalous.
 The regulated action, Eq. \ref{eq:Action_d_dim}, 
makes sense as long as we are expanding around a non-trivial 
gravitational background.
If we expand around a flat background, $R=0$, then the regulator  
is ill-defined. Alternatively we may use the regularization proposed in 
\cite{Englert,JMS}. In this case the regulator is well defined, as long
as we expand around a non-zero value of the classical scalar field.

\section{Standard Model with Local Scale Invariance}
We now consider the locally scale invariant extension of the
Standard Model \cite{ChengPRL}. The action in four dimensions may be
written as,
\ba
\mathcal{S} &=& \int d^4x \sqrt{-g}\Bigg[-{\beta\over 4} \mc H^\dag \mc H R'
+g^{\mu\nu} (D_\mu \mc H)^\dag(D_\nu \mc H) - \frac14
g^{\mu\nu}g^{\alpha\beta}(\mathcal{A}_{\mu\alpha} \mathcal{A}_{\nu\beta}
\nn\\
&+& \mathcal{B}_{\mu\alpha} \mathcal{B}_{\nu\beta})
 - {1\over 4}g^{\mu\rho}g^{\nu\sigma}\mathcal E_{\mu\nu}\mathcal E_{\rho\sigma} - \lambda    (\mc H^\dag \mc H)^2\Bigg].
\label{eq:S_EW}
\ea
Here $\mc H$ is the standard Higgs field doublet,
\[ \mc H(x) = \left( \begin{array}{c}
h_1(x) \\
h_2(x) \end{array} \right) .\]
The covariant derivative,
\ba
D_\mu = \partial_\mu -i g{\bf T\cdot A}_\mu - ig'{Y\over 2} B_\mu - fCS_\mu,
\ea
where $B_\mu$ is the $U(1)$ gauge field, ${\bf A}_\mu = \tau^aA^a_\mu$
is the $SU(2)$ gauge field multiplet and $S_\mu$ is the Weyl vector meson.
As usual $\bf T$ denotes the $SU(2)$ generators, $Y$ the $U(1)$ hypercharge
and $C$ the scaling or conformal charge. For the Higgs field $C=1$.
The field tensors $\mathcal{A}_{\mu\nu}$,
$\mathcal{B}_{\mu\nu}$ and $\mathcal{E}_{\mu\nu}$ are the
field strength tensors for the
gauge fields $A_\mu$, $B_\mu$ and $S_\mu$ respectively.
The scalar, $R'$, is related to the Ricci scalar, $R$, by the relationship,
\ba
R' = R - 6f S^\kappa_{\ ;\kappa} - 6f^2 g^{\mu\nu}S_\mu S_\nu.
\ea
In the action, Eq. \ref{eq:S_EW}, we have not included the spinor fields.
These can be easily added \cite{ChengPRL}.

The classical equations of motion admit a solution with the FRW scale
parameter $a(t)$ given by Eq. \ref{eq:a(t)} and the Higgs field,
\ba
\mc H_0^\dagger \mc H_0 = {3\beta\over 2\lambda} H_0^2
\label{eq:Higgs0}
\ea
and all the remaining fields equal to zero.
As before the Hubble parameter $H_0$ is constant in this case.
The Weyl vector field may also be nonzero depending on the initial
conditions. In this case also one may choose a gauge such that
$\mc H_0^\dagger \mc H_0 $ is constant \cite{AJS}.
In general, however, the Hubble parameter depends on time. As long as it 
varies slowly we can use the adiabatic approximation and ignore its 
time dependence while computing the Feynman diagrams. Similarly
we can treat the time dependence of the scale factor $a(t)$ in the 
adiabatic approximation \cite{JM09}.

The classical solution, Eq. \ref{eq:Higgs0}, breaks pseudo-scale invariance.
As we expand around this classical solution we find that the gravitational
constant is generated. The electroweak symmetry is broken and
the $W$ and $Z$ bosons acquire masses \cite{ChengPRL}. The $W$ boson mass
is found to be 
$M_W^2 = g^2 (\mc H_0^\dagger \mc H_0)$.
The Weyl meson also acquires a mass,
\ba
M_S^2 = \left(1+{3\beta\over 2}\right)2f^2 (\mc H_0^\dagger \mc H_0)\,,
\ea
due to the breakdown of pseudo-scale invariance. The Higgs boson disappears from the
particle spectrum and acts like the longitudinal mode of the Weyl vector meson
\cite{ChengPRL}. This phenomenon was illustrated in the context of
cosmological symmetry breaking in Ref. \cite{JMS}.
The fermions acquire masses
by their Yukawa interactions. The model predicts a cosmological
constant whose value can be adjusted by fixing $\lambda$ to fit the current
cosmological observations. Furthermore the Weyl meson acts like a dark
matter candidate \cite{ChengPRL,AJS}. Hence the model fits all the
particle and cosmological observations.
The classical solution for the Higgs field is related to the Planck Mass
by the formula,
\begin{equation}
\beta \left(\mc H_0^\dagger \mc H_0\right) = {M_{\rm PL}^2\over 4 \pi}.
\label{eq:MassPlanck}
\end{equation}
Furthermore the model predicts dark energy,
\begin{equation}
\rho_\Lambda =  \lambda\left(\mc H_0^\dagger \mc H_0\right)^2\ . 
\label{eq:DarkEnergy}
\end{equation}

The important issue that we need to settle is whether the model is
consistent quantum mechanically. This would be true as long as scale
invariance is not anomalous. Based on earlier works \cite{Englert,JMS,JM09}
and the arguments presented in the previous section,
we expect this to be true.
We next explicitly write down the regulated action in $d$ dimensions.
The action in $d$ dimensions
may be written as,
\ba
\mathcal{S} &=& \int d^dx \sqrt{-\bar g}\Bigg[-{\beta\over 4} \mc H^\dag \mc H 
\bar R'
+\bar g^{\mu\nu} (D_\mu \mc H)^\dag(D_\nu \mc H) - \frac14
\bar g^{\mu\nu}\bar g^{\alpha\beta}(\mathcal{A}_{\mu\alpha} \mathcal{A}_{\nu\beta}
\nn\\
&+& \mathcal{B}_{\mu\alpha} \mathcal{B}_{\nu\beta}) (\bar R'\,^2)^{-\epsilon/4}
 - {1\over 4}\bar g^{\mu\rho}\bar g^{\nu\sigma}\mathcal{E}_{\mu\nu}\mathcal{E}_{\rho\sigma}
(\bar R'\,^2)^{-\epsilon/4} 
 - \lambda    (\mc H^\dag \mc H)^2 (\bar R'\,^2)^{\, \epsilon/4}  \Bigg]\,,
\label{eq:S_EW_d}
\ea
where, as before, we denote all quantum gravity variables with a {\it bar}.
Here we have used the scalar $R'$, which transforms covariantly under
pseudo-scale transformations, in order to regulate the action. 
The regulated fermionic action, corresponding to the kinetic energy
terms, can also be written easily in $d$ dimensions as,
\ba
\mathcal{S}_{\rm fermions} =  \int d^dx \sqrt{-g} \Bigg[\bar{\Psi}\gamma^ce^\mu_ci
\Bigg(D_\mu-\frac12\sigma_{ab}e^{b\nu}
(D_\mu e^a_\nu - \Gamma^\rho_{\mu\nu}e^a_\rho)\Bigg)\Psi\Bigg],
\label{eq_S_fermion}
\ea
where $e^\mu_a$ is the tetrad. The scaling charge for fermion and the tetrad
$e^\mu_a$ are $c(d)$ and $b(d)/2$, respectively. In the connection $\Gamma^\rho_{\mu\nu}$ the derivatives of the metric are to be replaced by the suitable
pseudo-scale covariant derivatives with the suitable charge which can
be obtained
from Eq. \ref{eq:pseudo_ddim}.
The Yukawa interaction terms may be written in $d$ dimensions as,
\ba
\mathcal{S}_{\rm Yukawa} =  \int d^dx \sqrt{-\bar g} X_{ab}\bar{\Psi}_a {\mc H}\Psi_b
(\bar R'\,^2)^{\, \epsilon/8} + h.c.
\ea
for two fermion species $a$ and $b$. Here $X_{ab}$ are the Yukawa couplings.

The model is 
renormalizable as long as we ignore quantum gravity contributions. 
The gravitational sector of the model is expected to be 
non-renormalizable. It is very economical since it only introduces
only one vector field, the Weyl meson, besides the Standard Model fields.
The model
solves the hierarchy problem since it contains no physical scalar fields.
By hierarchy problem we refer to the stability of the electroweak symmetry
breaking scale to loop corrections. The problem arises primarily due to
the presence of physical scalar fields in the particle spectrum \cite{Martin}.
In the present model, all the scalar fields get eliminated and do
not appear as a physical particle, thereby solving the hierarchy problem.
The model predicts absence of the Higgs particle from the physical spectrum.
The Higgs particle essentially acts as the longitudinal mode of the
Weyl vector meson. 
As in the case of the Standard Model, the particle content of the model
is best seen by choosing a particular gauge, as discussed in 
\cite{ChengPRL,JMS}. This is similar to the unitary gauge in the 
Standard Model and all unphysical degrees of freedom do not appear
in this gauge. In this gauge the Higgs field does not appear in the 
Lagrangian and we recover the standard Einstein action besides the matter
terms. However for loop calculations it is convenient to use
a different gauge where the scalar fields appear as internal lines
but never appear as physical particles. Hence for loop calculations 
it may be better to work in a general gauge and then specialize to the 
unitary gauge at the end of the calculation.

The cosmological constant generated in this
model is expected
to be finite at all orders in perturbation theory due to scale invariance.
In Ref. \cite{JM09} we have explicitly demonstrated this at one loop order
in a toy model with a real scalar field.
In order to fit the small value of observed dark energy, we need to choose
a very small value of the coupling $\lambda$. The choice of small value
of the coupling is a shortcoming of the model. However once this is chosen
we do not expect fine-tuning at loop orders. The contribution due to the
Higgs field at higher orders in perturbation theory is extremely small since
the coupling is so small. Furthermore 
 we also need to include loop corrections due to the fermion
and vector fields. These are likely to be constrained due to scale invariance.
We see this as follows. If a field is minimally coupled to gravity
then its contribution to the Einstein equations arise through the 
energy-momentum tensor, $T_{\mu\nu}$. This is related to the conformal
current by the relationship
\begin{equation}
(J^\mu)_{;\mu} = T^\mu_\mu = 0\, .
\label{eq:traceEM}
\end{equation}
 The fact that the trace of the
energy momentum is zero as an operator identity 
clearly imposes constraints on the size of the 
vacuum energy that any field might contribute. Let us assume an isotropic and
homogeneous fluid such that the expectation
value of the trace, $<T^\mu_\mu>={\rm diag}(\rho,-p,-p,-p)$, where $\rho$
is the total energy density and $p$ the total pressure. We first consider a 
field which 
contributes only to relativistic energy density, with the equation
of state $p_{_R}=\rho_{_R}/3$, and to the vacuum energy density, with $p_{_V}
=-\rho_{_V}$.
Here $p_{_R}$ is the pressure due to the relativistic gas and $p_{_V}$ due to vacuum.
Similarly $\rho_{_R}$ is the energy density due to the relativistic gas and 
$\rho_{_V}$ due to vacuum.
From Eq. \ref{eq:traceEM} we obtain the constraint, $\rho-3p=0$, where
$\rho = \rho_{_R} + \rho_{_V}$ and $p=p_{_R}+p_{_V}$. In this case the constraint
implies that $\rho_{_V}=0$. Similarly if we assume that a field contributes 
only to non-relativistic and vacuum energy density, we find that 
$\rho_{_V} = -\rho_{_{NR}}/4$, where $\rho_{_{NR}}$ denotes the non-relativistic
energy density.
 
Hence we find that for a minimally coupled field the contribution to vacuum
energy is constrained and hence requires no fine tuning. Such a constraint
does not apply to the Higgs field since it does not couple minimally.  
 The loop contribution due to Higgs is small
as already explained. We shall explore the contribution due to Weyl meson 
in a separate publication. We expect it also to be contrained by scale
invariance. Hence we expect that no fine tuning may be required at loop 
orders. 

The main shortcoming
of the model is that
the parameters $\lambda$ and $1/\beta$ are found to be very small compared to
unity. The model provides no explanation for their extreme values. Furthermore
it is not clear how inflation arises in this model, although 
an inflationary solution to this model has also been suggested 
\cite{Kao}.
The model describes both the high energy physics
as well as the cosmological observations. Hence it is an attractive
generalization of the Standard Model of particle physics.

\section{Conclusions}
In this paper, following earlier works \cite{Englert,JMS,JM09},
 we have argued that scale invariance can be implemented
exactly in quantum field theory. We considered a locally scale invariant
extension of the Standard Model \cite{ChengPRL} and argued that it
fits both the particle physics and cosmological data. Furthermore
the model does not suffer from fine tuning problems due to absence of
scalar particles in the particle spectrum and due to scale invariance.
The cosmological constant
is expected to be finite at all orders in perturbation theory
and hence a prediction of the model. At one loop this was explicitly demonstrated in Ref. \cite{JM09}. The model is particularly attractive since it 
proposes only one extra field, namely the Weyl vector meson,
besides the Standard Model particles. This additional field also serves
as a dark matter candidate.
The model does contain some parameters which take extreme values. Both
the parameters $\lambda$ and $1/\beta$ are very small compared to unity.
So far we have no explanation for why they are so small. Furthermore
it is not clear how inflation arises in this model. 
The model
may be extended to include supersymmetry although it is not necessary for
the solution of the hierarchy problem. It may also be interesting
to explore grand unified models with local scale invariance.

\bigskip
\noindent
{\bf Acknowledgements:}  We 
thank K. Bhattacharya
 for collaborating during the initial stages 
of this project.

\end{document}